\documentclass[runningheads]{llncs}
\usepackage[T1]{fontenc}
\usepackage{graphicx}
\usepackage{color}
\usepackage{listings}
\usepackage{subcaption}
\usepackage{amsmath} 
\usepackage[inline]{enumitem}
\usepackage{import}
\usepackage{hyperref}
\usepackage{cleveref}
\usepackage{mytodonotes}
\usepackage[firstpage]{draftwatermark}
\definecolor{codegreen}{rgb}{0,0.6,0}
\definecolor{codegray}{rgb}{0.5,0.5,0.5}
\definecolor{codepurple}{rgb}{0.58,0,0.82}
\definecolor{backcolour}{rgb}{0.95,0.95,0.95}
\lstdefinestyle{pythonstyle}{
  language=Python,
  backgroundcolor=\color{backcolour},
  commentstyle=\color{codegreen}\itshape,
  keywordstyle=\color{blue}\bfseries,
  stringstyle=\color{codepurple},
  basicstyle=\ttfamily\scriptsize,
  breakatwhitespace=false,
  breaklines=true,
  captionpos=b,
  keepspaces=true,
  numbers=left,
  numbersep=5pt,
  numberstyle=\tiny\color{codegray},
  showspaces=false,
  showstringspaces=false,
  showtabs=false,
  tabsize=2,
  frame=single,
  xleftmargin=2em,
  framexleftmargin=1.5em,
  columns=fullflexible
}
\graphicspath{{images/}}
\begin{document}

\SetWatermarkText{
    \hspace*{3.7in}\raisebox{8.55in}{\href{https://eapls.org/pages/artifact_badges/}{\mbox{\includegraphics[width=1.4cm]{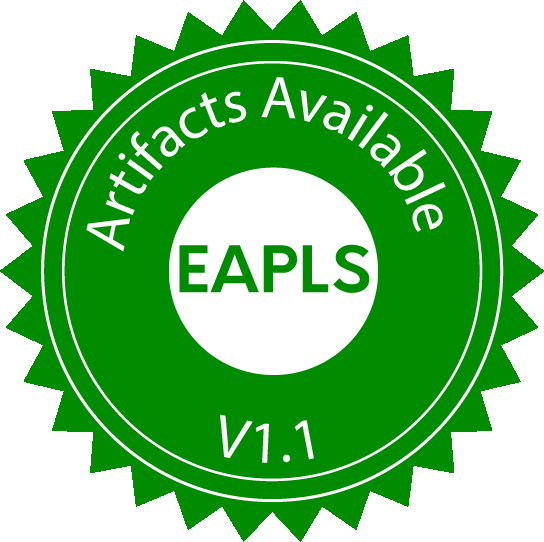}\includegraphics[width=1.4cm]{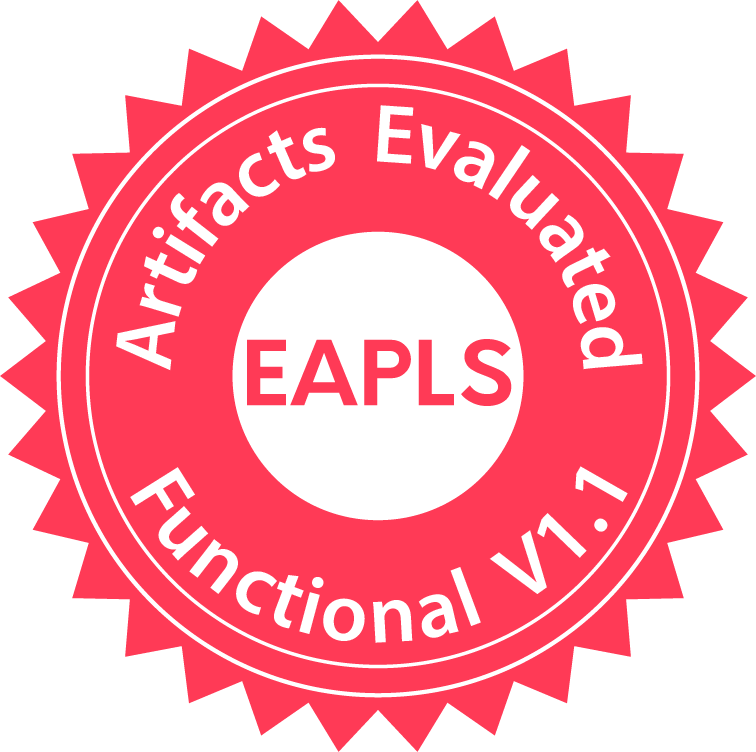}}}}} 
\SetWatermarkAngle{0}
\title{\textsc{Phyelds}: A Pythonic Framework for Aggregate Computing}
%
%
\author{
    Gianluca Aguzzi\orcidID{0000-0002-1553-4561} \and 
    Davide Domini\orcidID{0009-0006-8337-8990} \and 
    Nicolas Farabegoli\orcidID{0000-0002-7321-358X} \and 
    Mirko Viroli\orcidID{0000-0003-2702-5702}}
\authorrunning{G. Aguzzi et al.}
%
\institute{University of Bologna, Bologna, Italy\\
\email{\{gianluca.aguzzi, davide.domini, nicolas.farabegoli, mirko.viroli\}@unibo.it}}
\maketitle              

\begin{abstract}
\sloppypar
Aggregate programming is a field-based coordination paradigm 
 with over a decade of exploration and successful applications across domains including sensor networks, robotics, and IoT, 
 with implementations in various programming languages, such as Protelis, ScaFi (Scala), and FCPP (C++). 
A recent research direction integrates machine learning with aggregate computing, 
 aiming to support large-scale distributed learning and provide new abstractions for implementing learning algorithms. 
However, 
 existing implementations do not target data science practitioners, 
 who predominantly work in Python---the de facto language for data science and machine learning, 
 with a rich and mature ecosystem. 
Python also offers advantages for other use cases, 
 such as education and robotics (e.g., via ROS). 
To address this gap, 
 we present \textsc{Phyelds}, 
 a Python library for aggregate programming. 
\textsc{Phyelds} offers a fully featured yet lightweight implementation of the field calculus model of computation, 
 featuring a Pythonic API 
 and an architecture designed for seamless integration with Python's machine learning ecosystem. 
We describe the design and implementation of \textsc{Phyelds} 
 and illustrate its versatility across domains, 
 from well-known aggregate computing patterns 
 to federated learning coordination 
 and integration with a widely used multi-agent reinforcement learning simulator.
\end{abstract}
\section{Introduction}\label{sec:introduction}
The proliferation of distributed computing 
 systems---from sensor networks and IoT infrastructures to robotic swarms and edge-cloud continua---has 
 created an urgent need for programming abstractions 
 that can effectively manage collective behavior at scale~\cite{DBLP:journals/csur/Casadei23}.
Traditional approaches to distributed system development focus on individual devices,
 requiring programmers to manually orchestrate communication, 
 synchronization, 
 and coordination among potentially thousands of heterogeneous nodes.
This device-centric perspective becomes increasingly untenable as systems grow in scale and complexity, 
 motivating the emergence of \emph{macroprogramming} paradigms~\cite{10.1145/3712004}
 that shift the focus from individual devices to the collective behavior of the system as a whole.

Among macroprogramming approaches, 
 aggregate programming~\cite{DBLP:journals/computer/BealPV15} 
 has gained significant attention due to its principled foundation in the \emph{field calculus}~\cite{DBLP:journals/tocl/AudritoVDPB19}---a minimal functional language that models computation as the manipulation of \emph{computational fields}, 
 where values are distributed across space and evolve over time.
%
%
Application domains span sensor networks, swarm robotics, smart cities, and IoT systems, 
 where the ability to reason about global behavior while ensuring local execution is paramount.

A promising research direction integrates aggregate programming with machine learning~\cite{aguzzi2022machine}.
This synergy is bidirectional: 
 aggregate programming facilitates scalable distributed learning---e.g., in federated learning 
 and multi-agent reinforcement learning~\cite{DBLP:journals/iot/DominiFAVE26,aguzzi2023field,DBLP:journals/scp/DominiCAV24}---while machine learning enables data-driven adaptation within aggregate programs~\cite{aguzzi2022addressing}.
However, 
 realizing this potential faces a significant practical barrier: 
 existing implementations of aggregate programming---such as Protelis~\cite{DBLP:conf/sac/PianiniVB15}, 
 ScaFi~\cite{DBLP:journals/softx/CasadeiVAP22}, 
 and FCPP~\cite{DBLP:journals/scp/AudritoT24}---target languages (Java, Scala, and C++, respectively) 
 that are not commonly used by machine learning practitioners.
Python has become the \emph{de facto} standard for data science and machine learning, 
 boasting a rich ecosystem of libraries including TensorFlow~\cite{DBLP:conf/osdi/AbadiBCCDDDGIIK16}, 
 PyTorch~\cite{DBLP:conf/nips/PaszkeGMLBCKLGA19}, 
 and scikit-learn~\cite{DBLP:journals/jmlr/PedregosaVGMTGBPWDVPCBPD11}.
The absence of a Python-based aggregate programming framework thus hinders 
 the adoption of this paradigm in machine learning contexts 
 and limits the potential for cross-pollination between these research communities.

To bridge this gap, we present \textsc{Phyelds}\footnote{\url{https://github.com/phyelds/phyelds}}, 
 a lightweight and Pythonic library for aggregate programming.
\textsc{Phyelds} delivers a compact but comprehensive implementation of the field calculus, 
 offering an API that embraces Python's idioms and conventions.
Unlike its functional predecessors, 
 \textsc{Phyelds} adopts an imperative and object-oriented programming style 
 that is more familiar to the majority of Python developers, 
 lowering the barrier to entry for newcomers to aggregate programming.
The framework is designed with seamless integration in mind, 
 providing bindings for popular machine learning libraries 
 such as TensorFlow and PyTorch, 
 as well as multi-agent reinforcement learning environments like VMAS~\cite{bettini2022dars}.
We demonstrate its applicability through well-known aggregate computing patterns, 
 self-organizing federated learning~\cite{DBLP:journals/iot/DominiFAVE26,DBLP:journals/lmcs/DominiAEV26,DBLP:conf/ijcnn/DominiEACZLV25}, 
 and multi-agent reinforcement learning via integration with the VMAS simulator~\cite{bettini2022dars}, 
 where aggregate programming abstractions are leveraged to coordinate collective agent behavior.
 
Beyond machine learning, bringing aggregate programming into the Python ecosystem opens opportunities in other Python-dominated domains.
In \textbf{education}, Python's gentle learning curve makes it an ideal vehicle for introducing students to collective programming concepts.
In \textbf{robotics}, integration with ROS (Robot Operating System)~\cite{DBLP:journals/scirobotics/MacenskiFGLW22} enables the deployment of aggregate programs on physical robot swarms.

The remainder of this paper is structured as follows.
\Cref{sec:ac-overview} provides a brief overview of aggregate programming,
 covering the system model (\Cref{sec:system-model}),
 the field calculus operators (\Cref{sec:fc-operators}),
 and the alignment mechanism (\Cref{sec:alignment}).
\Cref{sec:phyelds} presents the design and implementation of \textsc{Phyelds}.
\Cref{sec:examples} illustrates the library through concrete examples,
 including common aggregate computing patterns,
 self-organizing federated learning,
 and integration with the VMAS multi-agent reinforcement learning simulator.
 Finally, \Cref{sec:conclusion} concludes with directions for future work.

\section{Aggregate Programming - A Brief Overview}\label{sec:ac-overview}

Aggregate computing (AC)~\cite{DBLP:journals/computer/BealPV15} is a macroprogramming paradigm 
 for the development of distributed systems, 
 where the focus is on the collective behavior of a system of devices, 
 rather than on the individual behavior of each device.
Starting from natural inspiration, 
 the computation is expressed as a manipulation over \emph{computational fields}, 
 where the value of a field at a given point in space and time represents the state of the device located there.
The field calculus~\cite{DBLP:journals/computer/BealPV15} is a minimal functional language 
 that captures the essence of aggregate programming, 
 providing a formal model for reasoning about field computations.
\subsection{System Model}\label{sec:system-model}

Typically,
an aggregate computing system consists of a collection of devices, 
each equipped with local computation capabilities, 
sensors for perceiving the environment, 
and communication interfaces for exchanging information with neighboring devices.
Each device is connected to a subset of other devices,
forming a dynamic and potentially unreliable network topology.
The communication model is typically based on message passing, 
where devices can send and receive messages asynchronously to and from their neighbors.

\begin{figure}
    \centering
    \def\svgwidth{0.8\textwidth}
\begingroup%
  \makeatletter%
  \providecommand\color[2][]{%
    \errmessage{(Inkscape) Color is used for the text in Inkscape, but the package 'color.sty' is not loaded}%
    \renewcommand\color[2][]{}%
  }%
  \providecommand\transparent[1]{%
    \errmessage{(Inkscape) Transparency is used (non-zero) for the text in Inkscape, but the package 'transparent.sty' is not loaded}%
    \renewcommand\transparent[1]{}%
  }%
  \providecommand\rotatebox[2]{#2}%
  \newcommand*\fsize{\dimexpr\f@size pt\relax}%
  \newcommand*\lineheight[1]{\fontsize{\fsize}{#1\fsize}\selectfont}%
  \ifx\svgwidth\undefined%
    \setlength{\unitlength}{640.40383647bp}%
    \ifx\svgscale\undefined%
      \relax%
    \else%
      \setlength{\unitlength}{\unitlength * \real{\svgscale}}%
    \fi%
  \else%
    \setlength{\unitlength}{\svgwidth}%
  \fi%
  \global\let\svgwidth\undefined%
  \global\let\svgscale\undefined%
  \makeatother%
  \begin{picture}(1,0.30150141)%
    \lineheight{1}%
    \setlength\tabcolsep{0pt}%
    \put(0,0){\includegraphics[width=\unitlength,page=1]{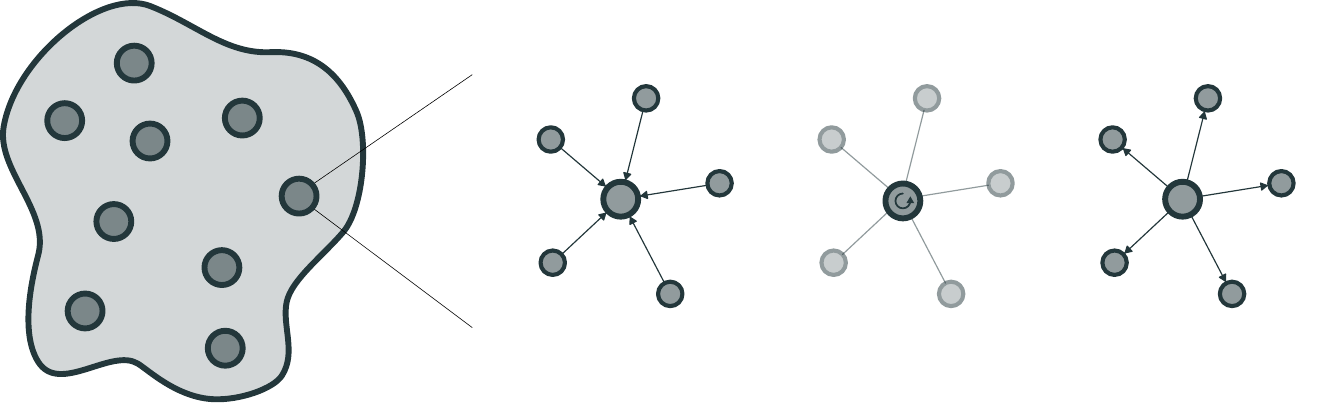}}%
    \put(0.46527123,0.02017){\color[rgb]{0.17254902,0.34117647,0.37254902}\makebox(0,0)[t]{\smash{\begin{tabular}[t]{c}\textbf{Sense}\end{tabular}}}}%
    \put(0.67662854,0.02268403){\color[rgb]{0.17254902,0.34117647,0.37254902}\makebox(0,0)[t]{\smash{\begin{tabular}[t]{c}\textbf{Compute}\end{tabular}}}}%
    \put(0.88618231,0.02027931){\color[rgb]{0.17254902,0.34117647,0.37254902}\makebox(0,0)[t]{\smash{\begin{tabular}[t]{c}\textbf{Interact}\end{tabular}}}}%
    \put(0,0){\includegraphics[width=\unitlength,page=2]{ac.pdf}}%
    \put(0.6867131,0.27571643){\color[rgb]{0.17254902,0.34117647,0.37254902}\makebox(0,0)[t]{\smash{\begin{tabular}[t]{c}computational round\end{tabular}}}}%
  \end{picture}%
\endgroup%

    \caption{Graphical representation of the aggregate computing system model.}
    \label{fig:ac-model}
\end{figure}

Each device executes the same aggregate program, 
which can be conceptually viewed as a function that takes the device's local state and the values received from neighbors as input,
and produces an output value that represents the device's contribution to the global computation.
Specifically,
each device organizes its execution into discrete rounds, 
where in each round it performs the following steps (\Cref{fig:ac-model}):
\begin{enumerate}
    \item \textbf{Sense}: the device collects the neighbors' messages, reads its sensors, and gathers the previous local state;
    \item \textbf{Compute}: the device executes the aggregate program using the collected information, producing a new local state and an output value;
    \item \textbf{Interact}: the device sends the output value to its neighbors, and stores the new local state for the next round.
\end{enumerate}

The rounds are not globally synchronized,
and they typically proceed asynchronously across devices.
This execution model is resilient to topology changes,
communication and device failures,
relying on an eventual behavior,
which is the foundational property of self-stabilization.

\subsection{Field Calculus Operators}\label{sec:fc-operators}
The field calculus provides a small set of core operators that enable the expression of complex distributed computations in a concise and composable manner.
In what follows, we briefly describe the main operators of the field calculus.

\begin{description}
    \item[rep] defines stateful computation among device rounds, allowing each device to maintain and evolve a local state.
        Evaluates an initial value during the first round and then updates it in subsequent rounds based on the previous state and an evolving function.
    \item[nbr] enables the exchange of information between neighboring devices, building a \emph{neighborhood field} that captures the values of neighboring devices aligned with that expression.
    \item[if] provides conditional branching, allowing devices to diverge in their execution paths based on local conditions.
        Only the subset of neighboring devices aligned with the same branch condition are visible inside the branch.
    \item[foldhood] aggregates information from neighboring devices using a specified binary operator, 
        enabling the computation of collective properties such as sums, counts, or maxima across the neighborhood.
\end{description}

On top of these core operators, 
 a rich library of higher-level constructs can be built, 
 including gradient computation, 
 information spreading and collection, 
 gossip protocols, 
 and leader election algorithms.

\subsection{Alignment Mechanisms}\label{sec:alignment}
The alignment is a fundamental aspect of aggregate programming,
ensuring that devices evaluating the same program are considered (aligned) and can correctly exchange information.
We refer to \emph{aligned devices} as those that satisfy the following two conditions:
\begin{enumerate*}[label=(\roman*)]
    \item the devices are executing the exact same program;
    \item the produced output (i.e., the value tree) resulting from the execution, reflects the same sequence of function calls and operator applications.
\end{enumerate*}
Any violation of these conditions results in \emph{misalignment}, 
where the devices are not considered part of the same computation.

When an aggregate program is executed,
two devices may be aligned for some parts of the program and misaligned for others.
Consider a program where an \texttt{if} operation is used.
Based on the boolean condition evaluated at runtime,
some devices may execute the \texttt{then} branch, 
while others execute the \texttt{else} branch.
This divergence at runtime leads to \emph{intended} misalignment,
where on the two branches,
only a subset of devices is aligned with each other.
%
Once the execution of the branches is completed,
all the devices will realign as they will converge to the same program point again.
The reconciliation of the devices is because of the assumption that all the devices execute the same program, 
and thus they will eventually converge to the same program point again,
despite the divergence at runtime.

In practice,
the alignment mechanism is implemented by associating a unique identifier to each aggregate operator which requires either a stateful operation (e.g., \texttt{rep}),
a neighbor interaction (e.g., \texttt{nbr})
or a network partitioning (e.g., \texttt{if}).
%

\section{The Phyelds Library}\label{sec:phyelds}
\textsc{Phyelds} was conceived with three key design goals.
First, 
we provide an \textbf{API} that is idiomatic to Python, 
 embracing its object-oriented and imperative programming style to lower the barrier to entry for developers.
First, existing frameworks are predominantly based on functional languages and 
 adopting an imperative style is crucial to make aggregate programming accessible to the majority of Python users.
Second, the library relies on a \textbf{lightweight and modular architecture} designed for applicability across diverse contexts, 
ranging from small-scale simulations and machine learning applications to deployment on physical devices.
To support this versatility, 
users can selectively incorporate only the components relevant to their specific use case.
Third, we ensure \textbf{integration with the Python ecosystem} 
by providing a simple execution framework for verifying the behavior of implemented blocks.
Additionally, we provide bindings to popular machine learning libraries, 
multi-agent reinforcement learning environments, and robotics frameworks.

\begin{figure}
    \centering
    \includegraphics[width=0.9\textwidth]{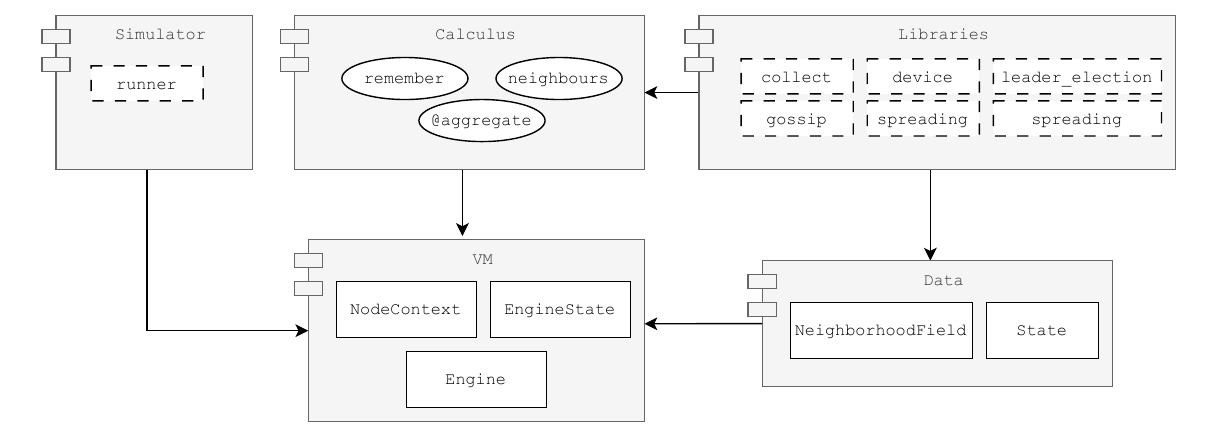}
    \caption{Architecture of \textsc{Phyelds}. Modules (components) contain classes (rectangles), submodules (dashed borders), and functions (ovals). Arrows indicate dependencies.}
    \label{fig:architecture}
\end{figure}

With these goals in mind,
the architecture of \textsc{Phyelds} is structured into distinct modules,
each responsible for a specific aspect of the aggregate programming model.
Namely, it comprises five principal components (\Cref{fig:architecture}): 
 the \texttt{VM}, the \texttt{Data}, 
the \texttt{Calculus}, the \texttt{Library}, and the \texttt{Simulator} modules.

\sloppypar
\subsubsection{VM Module.}
The VM module governs the execution of aggregate programs.
It provides the low-level APIs required to track device states,
manage function calls for the alignment mechanism,
and maintain execution contexts.
Specifically, 
this module handles gathering messages from neighbors,
persisting device state across rounds,
and orchestrating the execution of field calculus operators.
The principal classes in this module are \texttt{Engine}, \texttt{NodeContext}, and \texttt{EngineState}.
The \texttt{Engine} class embodies the core logic of the virtual machine
and serves as the interface between the VM and the rest of the library.
To use the engine, it must be initialized with
messages received from neighbors,
the previous state,
the device identifier,
and the device context (encapsulating time, sensor readings, and other contextual data).
Subsequently, the aggregate program is executed,
leveraging this information to perform computations
and generate a new state alongside messages for neighboring devices.
Finally, the engine is reset to prepare for the next round.
The typical usage pattern is illustrated in \Cref{lst:engine-usage}:
\begin{lstlisting}[style=pythonstyle, caption={Engine usage pattern.}, label={lst:engine-usage}]
from phyelds import engine
from phyelds.vm import Engine
# 1. Setup: initialize the engine with the current context
engine.get().setup(
    node_context=node_ctx,
    messages=received_messages,
    state=previous_state
)
# 2. Execute the aggregate program
result = my_aggregate_program()
# 3. Cooldown: reset engine and retrieve messages to send
new_state, messages_to_send = engine.get().cooldown()
\end{lstlisting}

\subsubsection{Data Module.}
The Data module contains the fundamental data structures representing computational fields.
The two principal components are the \texttt{State} and \texttt{NeighborhoodField} classes.
The \texttt{State} class maintains a device's state across rounds,
corresponding to the \texttt{rep} construct in traditional field calculus.
Implemented as a proxy object, it allows manipulation as a standard Python object.
Additionally, the \texttt{State} class stores the usage path to ensure the alignment mechanism functions correctly.
The \texttt{NeighborhoodField} class represents values received from neighboring devices,
corresponding to the \texttt{nbr} construct.
Although internally implemented as a dictionary mapping device identifiers to values,
this detail is abstracted from the user,
who interacts with it as a standard Python object.
The class supports operations such as element addition,
arithmetic operations between neighboring fields,
and field merging.
The class exposes a \texttt{foldhood} method that aggregates neighbor values using a caller-supplied binary operator
(e.g., a function like addition combining a current value (caller) with a neighbor value (supplier)).
Common aggregations such as \texttt{min} and \texttt{max} are provided as convenience methods 
 that are encoded via the \texttt{foldhood} operator.

\subsubsection{Calculus Module.}
The Calculus module provides the primary API for writing aggregate programs,
encompassing the core operators and constructs of the field calculus.
The central element is the \texttt{@aggregate} decorator,
which transforms a standard Python function into one executable within an aggregate context.
Annotated functions undergo a transformation as illustrated in \Cref{lst:aggregate-transform}:
\begin{lstlisting}[style=pythonstyle, caption={Transformation of an \texttt{@aggregate} function.}, label={lst:aggregate-transform}]
def aggregate(func):
    def wrapper(*args, **kwargs):
        # Enter the execution context
        engine.get().enter(func.__name__)
        # Apply code transformations if necessary
        result = transform_code(func)(*args, **kwargs)
        # Exit the execution context
        engine.get().exit()
        return result
    return wrapper
# User-defined aggregate program
@aggregate
def my_program(sensors):
    return sensors["temperature"] + 1
\end{lstlisting}
The \texttt{transform\_code} function is responsible for managing alignment. 
Under the hood, 
it parses the function's source code into an Abstract Syntax Tree (AST) 
and applies a node transformer to intercept \texttt{if} statements. 
To ensure the engine can track and match execution paths across devices, 
the transformer rewrites the code by wrapping the \texttt{if} and \texttt{else} bodies within explicit \texttt{align\_left()} and \texttt{align\_right()} context managers. 
\Cref{lst:ast-transform} shows an example of how this transformation is applied.
\begin{lstlisting}[style=pythonstyle, caption={Example of code transformation for conditionals.}, label={lst:ast-transform}]
# Original user code
if condition:
    do_something()
else:
    do_other()
# Transformed code executed by the engine
if condition:
    with align_left():
        do_something()
else:
    with align_right():
        do_other()
\end{lstlisting}

The \texttt{engine} object is a globally accessible instance containing the current VM.
It is initialized prior to program execution and reset after each round.
To support parallel execution (e.g., in simulation contexts),
this object is defined as a context variable (\Cref{lst:engine-global}):
\begin{lstlisting}[style=pythonstyle, caption={Engine as a context variable.}, label={lst:engine-global}]
from contextvars import ContextVar
from phyelds.vm import Engine
engine: ContextVar[Engine] = ContextVar("engine")
\end{lstlisting}
The other two fundamental operators are \texttt{remember} and \texttt{neighbors}, which implement the \texttt{rep} and \texttt{nbr} constructs of the field calculus.
The \texttt{remember} operator maintains state across rounds.
In standard field calculus, \texttt{rep} is a functional operator that takes an initial value and an update function:
\begin{lstlisting}[style=pythonstyle]
rep(0) { (value) => value + 1 }
\end{lstlisting}
While this style is native to functional languages, it is often less familiar to Python developers. 
Therefore, inspired by React's state management, calling \texttt{remember} returns a tuple containing a setter function and the current value, clearly separating state updates from value access:
\begin{lstlisting}[style=pythonstyle, caption={Using the \texttt{remember} construct.}, label={lst:remember-use}]
set_value, value = remember(0)
set_value(value + 1)  # Imperative state update
\end{lstlisting}
The \texttt{neighbors} construct facilitates interaction with neighboring devices.
It takes a value, transmits it to neighbors, and returns the values received from neighbors as a \texttt{NeighborhoodField}:
\begin{lstlisting}[style=pythonstyle, caption={Using the \texttt{neighbors} construct.}, label={lst:neighbor-use}]
nbr_values = neighbors(my_value)  # Returns NeighborhoodField
\end{lstlisting}
When the argument of the \texttt{neighbors} function is a \texttt{State} object, transmission occurs lazily:
the \texttt{State} value is only sent if it has changed since the previous round,
enabling efficient tracking of state updates.
\subsubsection{Library Module.}\label{sec:library-module}
The Library module builds upon the Calculus module,
providing a collection of higher-level constructs and building blocks for common aggregate programming patterns.
These blocks implement well-established algorithms from aggregate programming literature,
enabling developers to compose complex distributed behaviors from reusable components.

The module is organized into several sub-modules, each addressing a specific aspect of aggregate computation.
The \texttt{device} sub-module accesses device-specific information such as the local identifier (\texttt{local\_id}), spatial coordinates (\texttt{local\_position}), and sensor readings (\texttt{sense}).
The \texttt{time} sub-module offers temporal primitives including current simulation time, a round counter, and decay functions for time-dependent behaviors.
The \texttt{distances} sub-module computes inter-device distances using either Euclidean metrics based on positions or hop-count metrics for topology-based reasoning.
The \texttt{spreading} sub-module implements information dissemination patterns.
Its core function, \texttt{distance\_to}, computes a gradient (potential field) emanating from designated source nodes.
This serves as the foundation for higher-level operations such as \texttt{broadcast}, which propagates data outward from sources,
and \texttt{cast\_from}, which applies accumulation functions along the propagation path.
Complementing spreading, the \texttt{collect} sub-module implements convergecast patterns that aggregate information toward source nodes along gradient paths.
The \texttt{find\_parent} function establishes parent-child relationships based on potential values,
while \texttt{collect\_with} enables flexible aggregation using user-defined accumulation functions.
Specialized variants include \texttt{count\_nodes} for counting devices in sub-regions and \texttt{sum\_values} for numeric aggregation.
The \texttt{gossip} sub-module provides epidemic-style communication protocols for network-wide information sharing.
The generic \texttt{gossip} function spreads values across the network using a user-specified combination operator,
with convenience functions \texttt{gossip\_max} and \texttt{gossip\_min} for computing global extrema.
A self-stabilizing variant, \texttt{stabilizing\_gossip}, bounds propagation by the network diameter to ensure convergence.
Finally, the \texttt{leader\_election} module implements a well-known self-stabilizing algorithm~\cite{DBLP:conf/saso/MoBD18} via the \texttt{elect\_leaders} function, 
using random identifiers to break symmetry. 
It ensures that every device is within a specified distance $d$ from exactly one leader, 
 creating a stable Voronoi partition that converges regardless of the initial state. 
More details on the algorithm can be found in the original paper~\cite{DBLP:conf/saso/MoBD18} and 
 in further extensions~\cite{DBLP:journals/automatica/MoADB22,MO20203336}.

These building blocks can be composed to implement complex distributed algorithms.
For instance, combining leader election with gradient computation and collection enables hierarchical data aggregation systems
where elected leaders collect and summarize sensor readings from their respective regions.
Detailed usage examples are provided in \Cref{sec:examples}.

\subsubsection{Simulator Module.}
The Simulator module provides a framework for testing and debugging aggregate programs in a controlled environment.
It enables users to create virtual networks of devices, define their communication, and observe program execution over multiple rounds.

The module is built around three core abstractions.
The \texttt{Node} class represents an individual device characterized by a unique identifier, spatial position, and arbitrary sensor data.
The \texttt{Environment} class manages the collection of nodes and, crucially, determines the neighborhood topology via a configurable neighborhood function.
The \texttt{Simulator} class orchestrates execution using a discrete-event model, maintaining an event queue that schedules program execution across nodes.
Neighborhood definition is a key aspect of the simulator design.
Rather than hard-coding a specific topology, the module provides pluggable neighborhood functions to determine which nodes can communicate.
Built-in options include \texttt{radius\_neighborhood} (nodes within a specified Euclidean distance),
\texttt{k\_nearest\_neighbors} (each node connects to its $k$ closest peers), and \texttt{full\_neighborhood} (fully connected network).
Users can also define custom neighborhood functions to model arbitrary communication patterns or dynamic topologies.
The module also provides deployment utilities for generating common spatial distributions,
including regular grid layouts (\texttt{grid\_generation}), perturbed lattices (\texttt{deformed\_lattice}) for more realistic deployments, and random distributions within circular regions (\texttt{random\_in\_circle}).
For observability, the simulator supports \texttt{Monitor} objects attached to the simulation.
Monitors receive callbacks at the start, 
after each event, 
and at completion, enabling logging, visualization, 
or metric collection.

Additionally, the module integrates with the VMAS~\cite{bettini2022dars} framework through the \texttt{VmasEnvironment} class,
enabling aggregate programs to control agents in physics-based multi-agent reinforcement learning scenarios.
This integration allows researchers to combine aggregate programming abstractions with learned policies for coordinated robot control.

\section{Example of Use}\label{sec:examples}

In this section, we provide several examples of use of the \textsc{Phyelds} library\footnote{\url{https://github.com/domm99/artifact-coordination-2026-phyelds}}, 
 demonstrating how to implement common aggregate programming patterns 
 and how to integrate with machine learning frameworks for coordination-relevant learning scenarios.
\textsc{Phyelds} can be used in a normal Python environment, 
 such as a Jupyter notebook\footnote{Please, take a look at the online binder repository at \url{https://mybinder.org/v2/gh/phyelds/phyelds-examples/HEAD?urlpath=\%2Fdoc\%2Ftree\%2F\%2Fbinder\%2Fphyelds-example.ipynb}}, 
 and it also provides a native simulator for testing and debugging aggregate programs.

\subsubsection{Common Aggregate Computing Patterns.}\label{sec:ac-patterns}

A common and well-established pattern in AC is the construction of a \emph{channel} 
 between a source device and a target device. 
This pattern is frequently employed to model information flows, 
 logical routing structures, 
 or spatial constraints within a distributed system.

\begin{figure*}
    \centering
    \begin{subfigure}{0.33\linewidth}
        \centering
        \includegraphics[width=\textwidth]{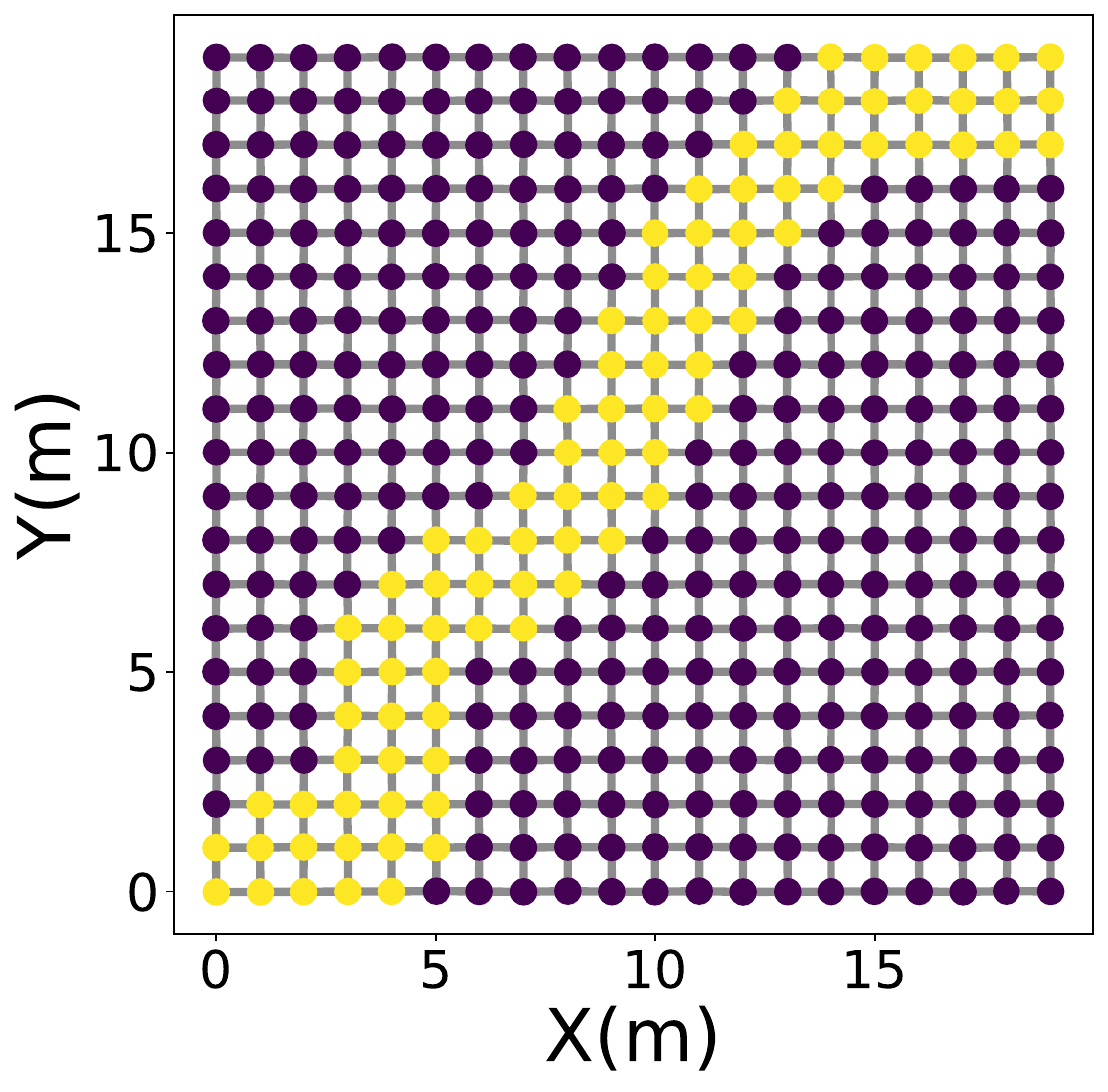}
        \caption{Channel example}
        \label{fig:channel}
    \end{subfigure}
        \begin{subfigure}{0.33\linewidth}
      \centering
      \includegraphics[width=\textwidth]{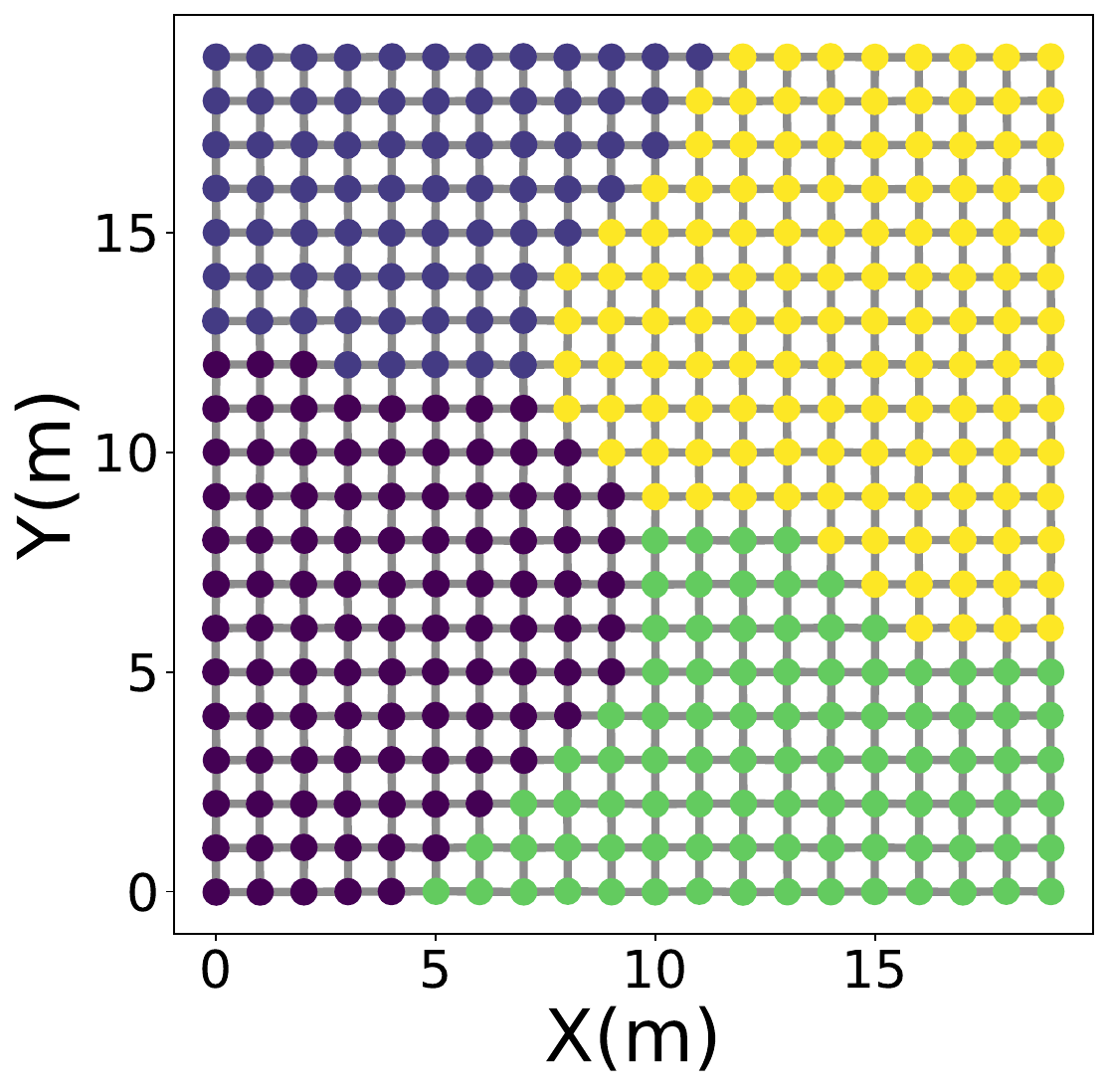}
      \caption{SCR example}
      \label{fig:scr}
    \end{subfigure}
    \caption{Examples of common aggregate programming patterns implemented in \textsc{Phyelds}.
    On the left, a channel is constructed between a source device (bottom left) and a target device (top right). The channel color is yellow. On the right, devices are partitioned into coordination regions (different colors).}
    \label{fig:ac-examples}
\end{figure*}

A channel (\Cref{fig:channel}) is defined as the set of devices that lie within 
 a given distance from a minimum path connecting the source and the target. 
This distance represents the \emph{width} of the channel, 
 allowing the model to range from a single-device path 
 to a wider region surrounding the optimal route. 
The final desired outcome is a boolean computational field, 
 where devices belonging to the channel evaluate to \texttt{True}, 
 while all others evaluate to \texttt{False}.

\Cref{lst:channel} shows the \textsc{Phyelds} code implementing this behavior. 
The first step consists of computing the computational field containing distances to neighboring devices, 
 which is obtained through the library function \texttt{neighbors\_distances} (line~\ref{line:channel-distances}). 
Based on this field, 
 the distance from the target device is then computed using the \texttt{distance\_to} operator (line~\ref{line:channel-distance-to}), 
  with the source argument given by the \texttt{``target''} sensor,
   which evaluates to \texttt{True} exclusively on the target device.
To construct the channel, 
 all devices lying on the minimum path between the source and the target are collected (line~\ref{line:channel-collect}). 
This is achieved by propagating information from the target distance field while constraining it with the \texttt{``source''} sensor, 
 effectively identifying the devices belonging to the shortest path (line~\ref{line:channel-shortest-path}). 
Finally, 
 to assign a channel width different from one (i.e., a single-device path), 
 each device computes its distance from the minimal channel and is included 
 in the final boolean field if this distance is smaller than the specified channel width (line~\ref{line:channel-width}).

\begin{lstlisting}[style=pythonstyle, caption={Construction of a channel from a source device to a target device.}, label={lst:channel}, escapeinside={(*@}{@*)}]
@aggregate
def main():
    distances = neighbors_distances()(*@\label{line:channel-distances}@*)
    target_distance = distance_to(sense("target"), distances)(*@\label{line:channel-distance-to}@*)
    nodes_in_path = collect_or(target_distance, sense("source"))(*@\label{line:channel-collect}@*)
    distance_from_path = distance_to(nodes_in_path, distances)(*@\label{line:channel-shortest-path}@*)
    channel = 1.0 if distance_from_path < width else 0.0 (*@\label{line:channel-width}@*)
    return channel 
\end{lstlisting}

\Cref{lst:channel-sim} shows how to set up and run the simulation for the channel example 
 (also available in the online Binder notebook).
A \texttt{Simulator} is first instantiated and configured with a radius-based neighborhood function (lines~\ref{line:sim-create}--\ref{line:sim-neighborhood}). 
A $20 \times 20$ deformed lattice topology is then created with positional noise to model a realistic device deployment (line~\ref{line:sim-lattice}). 
Each node is initialized with \texttt{source} and \texttt{target} sensor values set to \texttt{False}; 
 the first node in the list is designated as the source and the last as the target (lines~\ref{line:sim-sensor-start}--\ref{line:sim-sensor-end}). 
The aggregate program (i.e., the \texttt{main} function) is scheduled on every node with a fixed time step of $0.1$ seconds (lines~\ref{line:sim-schedule-start}--\ref{line:sim-schedule-end}). 
Finally, a \texttt{RenderMonitor} is attached to visualize the evolving field using colored nodes and drawn edges, 
 saving the result to a video file, and the simulation is run for $10$ seconds (lines~\ref{line:sim-render-start}--\ref{line:sim-run}).

\begin{lstlisting}[style=pythonstyle, caption={Simulation setup and execution for the channel example.}, label={lst:channel-sim}, escapeinside={(*@}{@*)}]
simulator = Simulator()(*@\label{line:sim-create}@*)
simulator.environment.set_neighborhood_function(radius_neighborhood(0.12))(*@\label{line:sim-neighborhood}@*)
deformed_lattice(simulator, 20, 20, 0.1, 0.01)(*@\label{line:sim-lattice}@*)
# initialise sensor data on each node
for node in simulator.environment.nodes.values():(*@\label{line:sim-sensor-start}@*)
    node.data = {"source": False, "target": False}
# designate source and target nodes
simulator.environment.node_list()[0].data["source"] = True
target = simulator.environment.node_list()[-1]
target.data["target"] = True(*@\label{line:sim-sensor-end}@*)
# schedule the aggregate program on every node
for node in simulator.environment.nodes.values():(*@\label{line:sim-schedule-start}@*)
    simulator.schedule_event(0.0, aggregate_program_runner, simulator, 0.1, node, main)(*@\label{line:sim-schedule-end}@*)
# configure rendering and run
RenderMonitor((*@\label{line:sim-render-start}@*)
    simulator,
    RenderConfig(
        effects=[DrawEdges(), DrawNodes(color_from="result")],
        mode=RenderMode.SHOW,
        save_as="channel.mp4",
        dt=0.1
    )
)
simulator.run(10)(*@\label{line:sim-run}@*)
\end{lstlisting}

A second widely used pattern in aggregate computing is known as 
 \emph{Self-Organizing Coordination Regions} (SCR)~\cite{DBLP:conf/coordination/CasadeiPVN19}. 
This pattern is inspired by classical results 
 in distributed systems~\cite{DBLP:conf/hicss/HeinzelmanCB00,DBLP:journals/trob/CortesMKB04,DBLP:journals/fgcs/PianiniCVN21}, 
 where large-scale systems are often partitioned into smaller subcomponents, 
 each governed by a special node---referred to as a \emph{leader}---responsible for coordination tasks.
SCR enables the spatial partitioning of devices into disjoint subregions, 
 effectively forming a Voronoi tessellation~\cite{DBLP:journals/csur/Aurenhammer91} of the space, 
 and the election of a leader within each region. 
Once leaders have been established, 
 each of them acts as the coordinator for its own region. 
In particular, 
 non-leader devices funnel their local information toward the leader through a convergecast process, 
 the leader computes a new result based on the collected data, 
 and subsequently disseminates the updated information back 
 to all devices in the region via a broadcast operation.

\Cref{lst:scr} illustrates an instance of this pattern, 
 where SCR is used to count the number of devices belonging to each region 
 and to propagate the resulting count to all devices within the same region. 
This is achieved by first electing leaders (line~\ref{line:scr-elect}), 
 then assigning devices to regions according to minimum-distance potentials(line~\ref{line:scr-potential}), 
 and finally combining convergecast (line~\ref{line:scr-count}) 
 and broadcast operations to aggregate (line~\ref{line:scr-broadcast})
 and disseminate information.

Although this example focuses on a simple aggregation task, 
 the SCR pattern is general and can be applied in significantly more complex scenarios. 
For instance, 
 it can be used to coordinate the training of a machine learning model independently within each region, 
 as discussed in the following subsection.

\begin{lstlisting}[style=pythonstyle, caption={Self-organizing Coordination Regions pattern.}, label={lst:scr},  escapeinside={(*@}{@*)}]
@aggregate
def main():
    distances = neighbors_distances()
    leader = elect_leaders(4, distances)(*@\label{line:scr-elect}@*)
    potential = distance_to(leader, distances)(*@\label{line:scr-potential}@*)
    nodes = count_nodes(potential)(*@\label{line:scr-count}@*)
    area_value = broadcast(leader, nodes, distances)(*@\label{line:scr-broadcast}@*)
    return area_value
\end{lstlisting}

\subsubsection{Self-Organizing Federated Learning.}\label{sec:sofl}
Federated Learning (FL)~\cite{DBLP:conf/aistats/McMahanMRHA17,DBLP:journals/corr/McMahanMRA16}
 is a distributed learning paradigm 
 in which multiple devices collaboratively train machine learning models 
 while keeping training data local and private.
Self-Organizing Federated Learning (SOFL)~\cite{DBLP:journals/lmcs/DominiAEV26,DBLP:journals/iot/DominiFAVE26}
 extends this paradigm by removing centralized coordination 
 and enabling devices to autonomously form multiple federations, 
 each learning a specialized model that better reflects local data characteristics. 
This is achieved by combining clustered federated learning 
 with self-organizing coordination mechanisms, 
 allowing the system to adapt dynamically to spatial deployment and 
 non-Identicaly and independently distributed (non-IID) data,
 specially when heterogeneity comes from spatial distribution 
 of partecipating devices~\cite{DBLP:journals/jors/DominiIALV26}.

\Cref{lst:fbfl} shows an implementation of a SOFL client using \textsc{Phyelds}. 
Each device interleaves local training with self-organizing coordination. 
The computation starts by maintaining local state through the \texttt{remember} construct (line~\ref{line:sofl-remember}), 
 storing the current model parameters and a logical training round counter. 
Each device then performs a local training step on its private dataset (line~\ref{line:local-train}), 
 producing an updated model and an associated training loss.
To enable self-organized federation formation, 
 devices compute a dynamic notion of distance based on model performance similarity. 
This is achieved through the function \texttt{loss\_based\_distances} (line~\ref{line:sofl-proxy}), 
 which derives a computational field reflecting how well neighboring models perform on local validation data. 
Leaders are then elected using \texttt{elect\_leaders} (line~\ref{line:sofl-elect}), 
 and a potential field toward each leader is computed via \texttt{distance\_to} (line~\ref{line:sofl-distance}), 
 implicitly defining coordination regions in a data-driven and adaptive manner.
Within each coordination region, 
 model updates are aggregated at the leader using a convergecast operation 
 implemented through \texttt{collect\_with} (line~\ref{line:sofl-collect}). 
In the example, 
 local models are combined and averaged to obtain a regional aggregated model (line~\ref{line:sofl-average}). 
The leader subsequently disseminates the updated model back to all devices 
 in the region using \texttt{broadcast}(line~\ref{line:sofl-distribute}), 
 ensuring that federation members are synchronized for the next training iteration (line~\ref{line:sofl-store}).

\begin{lstlisting}[style=pythonstyle, caption={Implementation of SOFL~\cite{DBLP:journals/iot/DominiFAVE26} in Phyelds.}, label={lst:fbfl},escapeinside={(*@}{@*)}]
@aggregate
def client(initial_model_params):
    set_value, value = remember((initial_model_params, 0))(*@\label{line:sofl-remember}@*)
    local_model, tick = value
    evolved_model, train_loss = local_training(local_model, training_data) (*@\label{line:local-train}@*)
    distances = loss_based_distances(evolved_model, validation_data)(*@\label{line:sofl-proxy}@*)
    leader = elect_leaders(threshold, distances)(*@\label{line:sofl-elect}@*)
    potential = distance_to(leader, distances)(*@\label{line:sofl-distance}@*)
    models = collect_with(potential, [evolved_model], lambda x, y: x + y)(*@\label{line:sofl-collect}@*)
    aggregated_model = average_weights(models)(*@\label{line:sofl-average}@*)
    area_model = broadcast(leader, aggregated_model, distances)(*@\label{line:sofl-distribute}@*)
    set_value((area_model, tick + 1))(*@\label{line:sofl-store}@*)
    return potential
\end{lstlisting}

\subsubsection{Integration with Third-Party Simulators.}\label{sec:vmas}

An additional relevant aspect of \textsc{Phyelds} is that, 
 although it provides a native simulator, 
 it is designed to seamlessly integrate with external simulation frameworks 
 by supplying appropriate bindings. 
This design enables \textsc{Phyelds} to operate in heterogeneous simulation ecosystems 
 without constraining the choice of the underlying simulator.

To demonstrate this capability, 
 we developed an integration with the VMAS simulator~\cite{bettini2022dars}, 
1 which is widely used in the context of multi-agent reinforcement learning (MARL) and swarm robotics.
Supporting an external simulator requires the implementation of two core components.
The first component is a simulator-specific \emph{runner} (in this case, a VMAS runner). 
At each simulated time step, 
 the runner retrieves the action computed by each agent, 
 executes these actions within the selected simulator, 
 and then provides the resulting observations or outcomes back to the agents.
The second component is an \emph{environment wrapper} for VMAS. 
This wrapper specifies how agents are initialized and 
 how their state is updated at each simulation step. 
It is invoked by the runner and acts as an abstraction layer 
 between \textsc{Phyelds} and the VMAS environment.

To showcase the functionality of the VMAS integration module, 
 we implemented a simple aggregate program that reproduces 
 flocking behavior according to the Vicsek model~\cite{vicsek1995novel}. 
Each agent $i$ is characterized by its position $\mathbf{r}_i(t)$ 
 and by the angle $\Theta_i(t)$ defining the direction of its velocity at time $t$. 
At each discrete time step $\Delta t$, 
 agents align their direction with that of their neighbors within a fixed interaction radius $r$, 
 up to a noise term:
\begin{equation}
\Theta_i(t+\Delta t)
=
\left\langle \Theta_j \right\rangle_{\lVert \mathbf{r}_i-\mathbf{r}_j\rVert < r}
+
\eta_i(t),
\end{equation}
where $\langle \Theta_j \rangle_{\lVert \mathbf{r}_i-\mathbf{r}_j\rVert < r}$ denotes 
 the average direction of all agents (including $i$) located within distance $r$ from agent $i$, 
 and $\eta_i(t)$ represents a stochastic noise term.

After updating its direction, each agent moves at constant speed $v$ according to
\begin{equation}
\mathbf{r}_i(t+\Delta t)
=
\mathbf{r}_i(t)
+
v\,\Delta t
\begin{pmatrix}
\cos \Theta_i(t)\\
\sin \Theta_i(t)
\end{pmatrix}.
\end{equation}

\begin{figure*}
    \centering
    \begin{subfigure}{0.29\linewidth}
        \centering
        \fbox{\includegraphics[width=\textwidth]{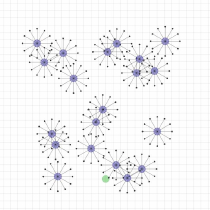}}
    \end{subfigure}
    \hfill
    \begin{subfigure}{0.29\linewidth}
      \centering
      \fbox{\includegraphics[width=\textwidth]{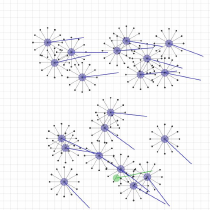}}
    \end{subfigure}
    \hfill
    \begin{subfigure}{0.29\linewidth}
    \centering
    \fbox{\includegraphics[width=\textwidth]{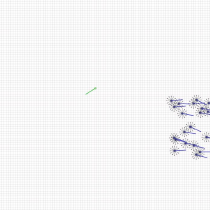}}
    \end{subfigure}
    \caption{Vicsek flocking simulation in VMAS using \textsc{Phyelds}. 
    Agents (cirlces) align their velocities over time (left to right), 
    exhibiting coherent collective motion.}
    \label{fig:vmas-example}
\end{figure*}

\Cref{lst:flocking} reports the aggregate program 
 implementing the Vicsek flocking model within the \textsc{Phyelds} framework. 
\Cref{fig:vmas-example}, instead, 
 shows the temporal evolution of the environment as obtained through 
 the native rendering capabilities of VMAS.
As time progresses (from left to right), 
 agents progressively align their directions, 
 eventually exhibiting coherent and collective motion.

\begin{lstlisting}[style=pythonstyle, caption={Implementation of the Vicsek Model in Phyelds using VMAS.}, label={lst:flocking}]
@aggregate
def action():
    myself = sense('agent')
    vel = myself.state.vel.squeeze()
    neighbors_info = neighbors(vel).exclude_self()
    velocities = [vel for vel in neighbors_info.values()]
    avg_vel = mean_velocity(velocities)
    theta = velocity_to_angle(vel, avg_vel)
    noise = perturbation()
    theta = theta + 0.1 * noise
    next_vel = [torch.cos(theta).item(), torch.sin(theta).item()]
    store("action", next_vel)
\end{lstlisting}

\section{Conclusion}\label{sec:conclusion}
This paper introduces \textsc{Phyelds}, 
 a Pythonic framework embodying a self-contained implementation 
 of the field calculus within the Python ecosystem. 
By combining the formal foundations of aggregate programming 
 with an imperative and object-oriented API,
 \textsc{Phyelds} lowers the barrier to adoption 
 for data science and machine learning practitioners.

Its modular architecture supports both core field calculus operators 
 and higher-level coordination patterns, 
 while enabling seamless integration with machine learning libraries 
 and external simulators such as VMAS. 
The presented examples demonstrate the expressiveness 
 of the framework across classical aggregate computing patterns, 
 self-organizing federated learning, 
 and multi-agent coordination scenarios.

Future work will focus on performance and scalability improvements, 
 deeper integration with mainstream ML toolchains, 
 systematic benchmarking against existing frameworks, 
 integration with other third-party simulators (e.g.,~\cite{DBLP:journals/corr/abs-2405-01562,DBLP:journals/jossw/HoevenKHPWRK25})
 and deployment in real-world edge and robotic systems (e.g.,~\cite{DBLP:conf/coordination/AguzziBBCCDFPV25,DBLP:conf/ccnc/AndruccioliCDFDPVV26})
 to further validate the practical potential of Python-based aggregate programming.

%
%
%
%
\bibliographystyle{splncs04}
\bibliography{bibliography}

\end{document}